\documentclass[aps,pra,preprint,groupedaddress]{revtex4}

\usepackage{graphicx}

\usepackage{amsmath,amsthm,amsfonts, latexsym}

\newcommand{\mtx}[2]{\left(\begin{array}{#1}#2\end{array}\right)}

\begin{document}


\title{Optimal Information Transfer and Real-Vector-Space Quantum Theory}


\author{William K. Wootters}
\affiliation{Department of Physics, Williams College, Williamstown, MA 01267, USA}


\date{8 January 2013}

\begin{abstract}
Consider a photon that has just emerged from a linear polarizing filter.  If the photon is then subjected to an orthogonal polarization measurement---e.g., horizontal vs vertical---the
photon's preparation cannot be fully expressed in the outcome: a binary outcome cannot reveal the value of a continuous variable.  However, a
stream of identically prepared photons can do much better.  To quantify this effect, one can compute the mutual information between 
the angle of polarization and the observed frequencies of occurrence of ``horizontal'' and ``vertical.''  Remarkably, one finds that the quantum-mechanical rule for
computing probabilities---Born's rule---{\em maximizes} this mutual information relative to other conceivable probability rules.  However, the maximization 
is achieved only because linear polarization can be modeled with a real state space; the argument fails when one considers the full set of complex states.  
This result generalizes to higher dimensional Hilbert spaces: in every case, one finds that information is transferred optimally from preparation to measurement
in the real-vector-space theory but not in the complex theory.  Attempts to modify the statement of the problem so as to see a similar optimization in the standard
complex theory are not successful (with one limited exception).  So it seems that this optimization should be regarded as a special feature of real-vector-space quantum theory.  

\end{abstract}


\maketitle


\section {Introduction}

In 1936 Birkhoff and von Neumann initiated an axiomatic approach to the foundations of quantum mechanics, taking as their starting point
postulates inspired by classical logic but adapted to the peculiar features of quantum theory \cite{Birkhoff}.  Though they showed that many characteristics of quantum theory could be captured in this way, they could also see that their logical approach would not
 lead uniquely to standard quantum theory.  In particular they noted that along with standard complex-vector-space quantum theory, the postulates could just as well be satisfied by
a theory based on a real or quaternionic Hilbert space \footnote{In fact, from this logical starting point, it has turned out to be difficult even to narrow the set of possibilities to just these
three theories---real, complex, and quaternionic.  Major theorems along these lines can be found in Refs.~\cite{Piron, Soler}; see also Ref.~\cite{Holland}.  Still, nothing in these papers favors the complex theory over the real or quaternionic theory.}.

Over the years other authors have taken other approaches to axiomatization and have found reasonable assumptions that favor the complex theory over the 
real and quaternionic models.  One successful strategy along these lines has been to insist on 
the existence of an uncertainty 
principle of a specific form \cite{Stueckelberg0, Stueckelberg1, Lahti}.  Another approach put forward by several authors relies on the fact that in standard quantum theory,
it is possible to carry out a complete tomographic reconstruction of the state of a multipartite system entirely by means
of local measurements on the individual components (taking into account correlations), with no need for global measurements on pairs of subsystems \cite{Araki, Bergia, Wootters1, Mermin, Hardy, Barrett, Chiribella}.  The real-vector-space theory does
not have this property; so by adopting local tomography as an axiom, one rules out the real case.  Surely, 
though, much of the appeal of these arguments comes from the fact that they succeed in leading us to what we believe to be the 
correct answer.  If we had found ourselves living in a world that seemed to be well described by real-vector-space
quantum theory, we would not have regarded it as a logical problem that tomography requires global measurements.  It would simply be another peculiar feature of quantum theory, like entanglement \footnote{At least in real-vector-space quantum theory, one never needs to make global measurements involving more than
{\em two} subsystems \cite{HardyWootters}.}.  (I admit, though, that the local tomographic property of the complex theory does feel as if it could be a clue to something deeper.)   

In this paper I would like to point out a particular property of real-vector-space quantum theory that I find
especially intriguing: the transfer of information from a preparation to a measurement is {\em optimal} (in a sense to be explained below).  Standard quantum theory does not
have this property.  So if we were trying to find a simple set of axioms that would generate real-vector-space quantum theory, we might well
find ourselves adopting optimal information transfer as one of our axioms.  This property of the real theory has been known for years---it appears in my
1980 doctoral dissertation \cite{Wootters2}---but I would like to give a somewhat simpler and more intuitive presentation of it here.   

One motivation for studying real-vector-space quantum theory is simply to shed light on the standard theory by comparison.  But I would also
like to keep open the possibility that the real-vector-space theory might turn out to be of value in its own right for describing our world.
Several authors have given us reasons
for not discounting this possibility.  In a series of papers published around 1960, Stueckelberg and his collaborators developed an alternative formulation of 
quantum field theory based on a real Hilbert space \cite{Stueckelberg1, Stueckelberg2, Stueckelberg3}.  In order to allow the existence of 
an uncertainty principle, Stueckelberg imposes a specific restriction on all the observables of the real-vector-space theory: every observable is required to commute with a certain operator that we can write as $I \otimes J$, where $J$ is the $2 \times 2$ matrix
$\mtx{cc}{0 & -1 \\ 1 & 0}$ and $I$ is the identity operator.  (In the context of Stueckelberg's papers $I$ is the identity on an infinite-dimensional real Hilbert space.)  In effect, this restriction forces the matrix representing any observable to be composed of $2 \times 2$ blocks of the form
$
\mtx{cc}{a & -b \\ b & a}.
$
Such $2\times 2$ blocks add and multiply like complex numbers; so the theory becomes equivalent to the usual complex theory.
One of the points Stueckelberg and his collaborators make in these papers is that in this formulation the time-reversal
operator becomes linear, rather than antilinear as in the complex formulation.  Around the same time, Dyson made the same point and argued that by bringing the time-reversal
operator into our formalism, we are in effect basing our quantum theory on the field of real numbers \cite{Dyson}.  

More recently Gibbons and his collaborators have argued that the complex structure in quantum theory is intimately related to the classical idea of time, and that both time itself and the associated complex structure could prove to be emergent features \cite{Gibbons1,Gibbons2}. 
In other work, Myrheim has pointed out that if one wants a version of the canonical commutation relation $[x,p] = i\hbar$ in a discrete
system with finitely many values of position and momentum, one cannot use standard complex quantum theory: the trace of any commutator is zero
in a finite-dimensional space, but the trace of $i\hbar$ is not zero.  On the other hand, if we replace $i\hbar$ with $J\hbar$ (the same $J$ as
above), both sides of the equation have zero trace and there is no contradiction \cite{Myrheim}.  In the present paper I do not particularly build on any of 
these observations except insofar as they suggest that a real-vector-space version of quantum theory
might be used to describe our actual world, and that the theory is worth studying for this reason as well as for whatever insights it might provide about 
standard quantum mechanics.

I begin in Section II by saying what I mean by ``real-vector-space quantum theory." Then in Sections III and IV I present the property 
of optimal information transfer, first for a two-dimensional state space and then in $d$ dimensions.  As I have said, standard complex quantum theory does not have this 
property, and it is interesting to ask whether a revised statement of the problem might yield a positive answer even in the complex case.  This is the subject of Section V.  Section VI then summarizes our findings.  

\section{Real-Vector-Space Quantum Theory}

One can summarize the basic structure of standard quantum theory in the following four statements:
\begin{enumerate}
\item A pure state is represented by a unit vector in a Hilbert space over the complex numbers.
\item An ideal repeatable measurement is represented by a set of orthogonal projection operators whose supports span the vector
space.  When a state $|s\rangle$ is subjected to the measurement $\{P_1, \cdots, P_m\}$, the probability of the 
$i$th outcome is $\langle s |P_i|s\rangle$.  When the $i$th outcome occurs, the system is left in a state
proportional to $P_i|s\rangle$.   
\item A reversible transformation is represented by a unitary operator $U$.  That is, for any initial state $|s\rangle$, the operation takes $|s\rangle$ to $U|s\rangle$.
\item A composite system has as its state space
the tensor product of the state spaces of its components.  
\end{enumerate}
Of course other states, measurements and transformations are possible.  Mixed states are averages of projection operators on pure states, and there also exist non-orthogonal
measurements and irreversible transformations.  But all such generalizations can be obtained from the 
cases listed above by applying them to a larger system and possibly discarding part of the system.  I have chosen the above
formulation partly to keep the discussion simple, but also because I do tend to think of orthogonal measurements and pure states as being more fundamental than their generalizations.  

The real-vector-space theory has essentially the same structure, except that all vectors and matrices are limited to real 
components.  The only changes in the above list are that ``complex'' is to be replaced by ``real'' in item 1, and ``unitary'' is to be replaced 
by ``orthogonal'' in item 3.  

One might wonder what the analogue of the Schr\"odinger equation is in the real-vector-space theory.  The Schr\"odinger
equation generates a unitary transformation through a Hermitian operator, the Hamiltonian:
\begin{equation}
i\hbar \frac{d}{dt} |s\rangle = H|s\rangle.
\end{equation}
If $H$ is time independent, the unitary operator it generates over a time $t$ is $U(t) = e^{-iHt/\hbar}$, since $|s(t)\rangle 
= U(t)|s(0)\rangle$ solves the differential equation.  The analogous equation in the real-vector-space case should have
an antisymmetric real matrix in place of $-iH$, since such a matrix generates orthogonal transformations.
We can write the differential equation as
\begin{equation}
\frac{d}{dt} |s\rangle = S |s\rangle,  \label{Stueckelberg}
\end{equation}
where $S$ is an antisymmetric real operator.  I like to call $S$ the ``Stueckelbergian'' in honor of Ernst Stueckelberg (who 
of course did not use this term).  If $S$ is time independent, then the general
solution of Eq.~(\ref{Stueckelberg}) is 
$|s(t)\rangle = e^{St}|s(0)\rangle$.

Another reasonable question is whether, for example, in a two-dimensional real space the operator 
\begin{equation}  \label{R}
R = \mtx{cc}{1 & 0 \\ 0 & -1}
\end{equation}
should be allowed to count as a possible transformation \cite{Aaronson}.  It is an orthogonal matrix, so according to the above rules it does count.  But there is
no $2 \times 2$ Stueckelbergian that can generate this operator.  This is because the operator $R$ represents a 
reflection, not a rotation, and there is no continuous set of orthogonal transformations on a two-dimensional real space
that takes us from the identity operator to a reflection operator.  

Nevertheless, in a real-vector-space world it would still be possible to realize the operation $R$ continuously 
by bringing in an ancillary two-dimensional system (that is, an ancillary ``rebit'').  To effect the transformation
\begin{equation}
\mtx{c}{s_1 \\ s_2} \rightarrow \mtx{c}{s_1 \\ -s_2},
\end{equation}
we can perform a controlled rotation on our ancillary rebit, conditioned on the state of the original rebit.  
By rotating the ancillary rebit by half a complete cycle, we can pick up the desired factor of $-1$.  So it seems reasonable to 
allow orthogonal matrices with negative determinant to count as possible transformations.

\section{Optimal Transfer of Information: The Two-Dimensional Case}

Consider the following simple scenario.  A stream of photons emerges from a linearly polarizing filter with its preferred axis oriented at an
angle $\theta$ from the horizontal.  Somewhere further along the photons' path there is a polarizing beam splitter and a pair of single-photon detectors, which together force each photon to yield either the horizontal outcome or the vertical outcome.  The probability of ``horizontal'' is $p_{0}(\theta) = \cos^2\theta$.  (The subscript ``0'' distinguishes this function from other hypothetical functions to be considered shortly.)  This function allows someone observing the measurement results to gain information about the angle $\theta$. 

This scenario illustrates a typical feature of a quantum measurement: a measurement on a single instance of a system (in this case a single photon) cannot convey complete information about the system's preparation.  But a large statistical sample of measurements on identically prepared copies can eventually home in on the values of the preparation parameters (in this case the single parameter $\theta$).  One does not encounter this limitation in classical physics, at least not for pure states: if a particle is placed at position $x$ with momentum $p$, a measurement can directly reveal those values.  This difference between classical and quantum physics reflects the fact that quantum theory is inherently probabilistic.  

In our specific example, one can ask how well the information about 
$\theta$ is conveyed through the observed results.  Specifically, one can quantify the mutual information between the measurement
results and the value of $\theta$.  As we will see shortly, given the
limitation imposed by the probabilistic nature of the polarization measurement, the transfer of information is optimal in the limit of a large number of trials.   
That is, in this example anyway, quantum mechanics orchestrates the optimal conveyance of information from the preparation to the
measurement outcome.  

Before justifying this statement, I want to note the sense in which we are effectively framing the problem in the real-vector-space theory.
By limiting the possibilities to linear polarizations, we are ruling out all the polarization states one would normally represent with 
vectors having a nonzero imaginary part (circular and elliptical polarizations).  We will return to this point toward the end of this section.

Now let us make the statement precise.  We do so by comparing our actual world, in which the probability of ``horizontal'' is
$p_{0}(\theta) = \cos^2\theta$, to a fictitious world in which the probability is given by some arbitrary function $p(\theta)$.  
In such a world, 
let $N$ photons, each prepared with linear polarization angle $\theta$, be subjected to a horizontal-vs-vertical polarization measurement.  
Let $n$ be the number of these photons that yield the outcome ``horizontal.''  The mutual information between the measurement results and the value of $\theta$ is based on the Shannon entropy $H$ and is defined to be
\begin{equation}  \label{mutualinfo}
I(n:\theta) = H(n) - H(n|\theta) = -\sum_{n=0}^N P(n)\ln P(n) +\frac{1}{2\pi} \int_0^{2\pi}\left( \sum_{n=0}^N P(n|\theta) \ln P(n|\theta)\right) d\theta.
\end{equation}
Here we have assumed a uniform {\em a priori} distribution of $\theta$ over the interval $[0, 2\pi]$.  (This is a crucial assumption that we
discuss further below.)  $P(n|\theta)$ is the probability
of getting the horizontal outcome exactly $n$ times if the photons are prepared in the state $\theta$, and $P(n)$ is the probability of
getting the horizontal outcome exactly $n$ times in the absence of any information about $\theta$ (that is, when $\theta$ is uniformly distributed).  Both $P(n|\theta)$ and $P(n)$ depend on the function $p(\theta)$.  Note that in Eq.~(\ref{mutualinfo}) we have written the mutual information as the average amount of information gained about the
integer $n$ upon learning the value of $\theta$.  It does have this interpretation, but it can alternatively be interpreted
as the average amount of information one gains about the value of $\theta$ upon learning the value of $n$.  (Mutual information is
symmetric in its two arguments.)  This latter interpretation
is more descriptive of the scenario we are imagining, in which an observer at the polarizing beam splitter is trying to 
learn about the value of $\theta$.  

It turns out that for large $N$, $I(n: \theta)$ grows as $(1/2)\ln N$.  We therefore consider the following limit, which has a finite
upper bound:
\begin{equation}
\tilde{I} = \lim_{N \rightarrow \infty} \left[ I(n:\theta) - \frac{1}{2}\ln N \right].
\end{equation}
We want to show that of all conceivable probability functions $p(\theta)$, the quantum mechanical
function $p_0(\theta) = \cos^2\theta$ gives $\tilde{I}$ its largest possible value.  

At this point we could proceed to compute $\tilde{I}$ starting from Eq.~(\ref{mutualinfo}), but the calculation will be simpler, and I hope clearer, if we 
abstract the problem away from its quantum mechanical setting.  The important point to notice is that
$I(n: \theta)$ depends on the probability function $p(\theta)$ only through the {\em measure} it induces on the binary 
probability space.  That is, before we have any knowledge of $\theta$, we can use $p(\theta)$
and the assumed uniform distribution of $\theta$ to figure out how likely it is that the probability of ``horizontal'' lies in any 
given interval, and it is this weighting function that figures into $I(n: \theta)$.   

The more abstract problem, then, can be stated as follows.  Consider a two-outcome probabilistic experiment, and let $(p_1,p_2)$ denote a point
in the 
binary probability space with $p_1$ corresponding to outcome \#1.  The experiment is run $N$ times, and outcome {\#}1 is observed to occur $n$ times.  This observation gives the experimenter information
about $(p_1, p_2)$ \footnote{The experimenter is trying to determine the value of an unknown probability.  It may seem that this 
problem cannot be framed except in the context of an objective interpretation of the concept of probability, but this is not the case.
The representation theorem of de Finetti shows how to express this kind of question within a subjective interpretation \cite{deFinetti}.  We note
that the {\em quantum} de Finetti theorem does not hold in real-vector-space quantum theory \cite{Caves2}, but this fact does not
preclude a subjective interpretation of probability in our problem.  In our problem the experimenter is trying to refine a distribution over ordinary probability space, to which the classical de Finetti theorem applies.  
}.  The mutual information $I$ between the value 
of $n$ and the value of $(p_1, p_2)$ depends on the experimenter's {\em a priori} measure on probability space.  Our problem is to
find the {\em a priori} measure that maximizes
the limit 
\begin{equation} \label{newtilde}
\tilde{I} = \lim_{N\rightarrow\infty} \left[I - \frac{1}{2}\ln N\right].
\end{equation}
(The optimal measure will turn out to be unique.)  We want to show that this optimal measure is the one induced
by the quantum probability function $p_0(\theta) = \cos^2\theta$ when $\theta$ is uniformly distributed.

\begin{figure}[h]
\begin{center}
\includegraphics[scale = 0.85]{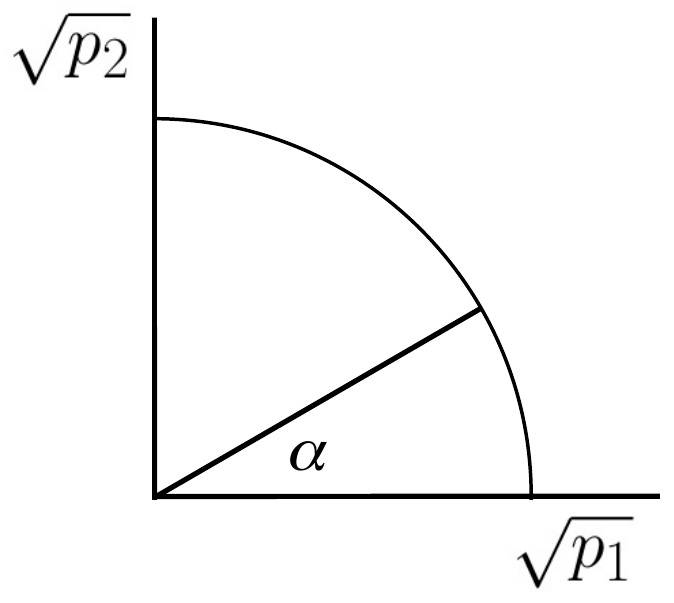}
\end{center}
\vspace{-10mm}
\caption{The relation between $(p_1,p_2)$ and $\alpha$.}
\label{rootfigurenew}
\end{figure}

In order to tackle this problem we need to choose a parameterization of the binary probability space.  We could use $p_1$ or $p_2$ as our 
parameter, but it turns out to be more convenient to use a different parameter $\alpha$ defined by
$(p_1, p_2) = (\cos^2 \alpha, \sin^2 \alpha)$, where $0 \le \alpha \le \pi/2$.  The relation between $\alpha$ and $(p_1, p_2)$ is illustrated in Fig.~\ref{rootfigurenew}. 
(One might object that we seem to be smuggling some quantum mechanics into the 
calculation here, but we are not.  The results will be entirely independent of our choice of parameter.  Our choice merely simplifies the
calculation.)  Let $K(\alpha)d\alpha$ be the {\em a priori} measure on the set of 
values of $\alpha$, 
normalized so that $\int_0^{\pi/2} K(\alpha) d\alpha = 1$.  The mutual information between $\alpha$ and $n$ can be 
written as
\begin{equation} \label{Kinfo}
I(\alpha : n) = h(\alpha) - h(\alpha| n) = - \int_0^{\pi/2} K(\alpha) \ln K(\alpha) d\alpha + 
\sum_{n=0}^N P(n) \int_0^{\pi/2} P(\alpha| n) \ln P(\alpha | n) d\alpha,
\end{equation}
where $h(\alpha)$ and $h(\alpha |n)$ are differential entropies \footnote{The differential entropy is not the limit of the entropy of a discretized version of the 
continuous variable.  However, a {\em mutual information} involving a continuous 
variable, being the {\em difference} between two differential entropies, is indeed the limit of the discretized mutual information \cite{Cover}.}.  
Here $P(\alpha | n)$ is the probability distribution the experimenter assigns to $\alpha$ after seeing the value $n$, and 
$P(n)$ is the {\em a priori} probability of the value $n$ as computed from the distribution $K(\alpha)$.  (Note that if $K(\alpha)$ is derived from the probability function $p(\theta)$ under the assumption that $\theta$ is uniformly distributed, then
$I(\alpha : n)$ is exactly equal to the quantity $I(n : \theta)$ given in Eq.~(\ref{mutualinfo}).)
The point of the next paragraph is to show that under modest assumptions about the function $K(\alpha)$, in the limit of very large $N$ the second term on the right-hand
side of Eq.~(\ref{Kinfo}) becomes {\em independent} of $K(\alpha)$.  So we will only have to think about maximizing the first term.  

To evaluate this second term, we need to write down expressions for $P(n)$ and $P(\alpha | n)$.  
We have
\begin{equation}  \label{PandP1}
P(n) = \int_0^{\pi/2} P(n|\alpha) K(\alpha) d\alpha \hspace{1cm} 
\end{equation}
and
\begin{equation}  \label{PandP2}
P(\alpha | n) = \frac{P(n|\alpha) K(\alpha)}{P(n)},
\end{equation}
where $P(n|\alpha)$ is given by the binomial distribution:
\begin{equation}
P(n|\alpha) = \frac{N!}{n!(N-n)!}\, p_1^n p_2^{N-n}
\end{equation}
with $p_1 = \cos^2\alpha$ and $p_2 = \sin^2\alpha$.  For any value of $p_1$ strictly between $0$ and $1$, it is possible 
to choose $N$ large enough that the binomial distribution is well approximated by a Gaussian:
\begin{equation} \label{gaussian}
P(n|\alpha) \approx \frac{1}{\sqrt{2\pi N p_1p_2}}\exp\left[ - \frac{N}{2p_1p_2}\left(n/N - p_1\right)^2 \right].
\end{equation}
As $N$ gets very large this distribution, regarded as a function of $n/N$, becomes arbitrarily highly peaked around $n/N = p_1$.
Let $\alpha^{(n)}$ be defined so that $n/N = \cos^2\alpha^{(n)}$.  That is, $\alpha^{(n)}$ is the value of $\alpha$ corresponding to
the observed outcome $n$.  Then
in the above exponent, we can approximate
the quantity $(n/N-p_1)$ as
\begin{equation}
(n/N-p_1) = \cos^2\alpha^{(n)} - \cos^2\alpha \approx \frac{d\left(\cos^2\alpha\right)}{d\alpha} \Delta \alpha = (-2\cos\alpha\sin\alpha) \Delta\alpha
=-2\sqrt{p_1p_2}\,\Delta\alpha,
\end{equation}
where $\Delta\alpha = \alpha^{(n)} - \alpha$.  This gives us
\begin{equation}  \label{14}
P(n|\alpha) \approx \frac{1}{\sqrt{2\pi N p_1p_2}} \exp\left[-2N(\Delta\alpha)^2\right].
\end{equation}
Inserting this expression into Eq.~(\ref{PandP1}), we again use the fact that the Gaussian is very highly
peaked so that we can (i) extend the integral from $-\infty$ to $\infty$ without changing its value appreciably and (ii) evaluate everything outside the exponential at $\alpha = \alpha^{(n)}$.  Then we get
\begin{equation} \label{Pnapprox}
P(n) \approx  \frac{K\hspace{-1mm}\left(\alpha^{(n)}\right)}{2N\cos\alpha^{(n)}\sin\alpha^{(n)}}.
\end{equation}
We now use Eqs.~(\ref{PandP2}), (\ref{14}) and (\ref{Pnapprox}) to approximate $P(\alpha | n)$:
\begin{equation}
P(\alpha | n) \approx \sqrt{\frac{2N}{\pi}}\,\exp\left[-2N(\Delta\alpha)^2\right].
\end{equation}
Using this expression and again relying on the narrowness of the Gaussian, we get
\begin{equation}
\int_0^{\pi/2} P(\alpha| n) \ln P(\alpha | n) d\alpha \approx  \frac{1}{2}\ln\left(\frac{2N}{\pi e}\right).
\end{equation}
Since this expression does not depend at all on $n$, it factors out of the sum in Eq.~(\ref{Kinfo}), so that the only sum we have to
do is $\sum_n P(n)$, which is unity by definition.  Putting the pieces together, we arrive at 
\begin{equation}
I(\alpha : n) \approx - \int_0^{\pi/2} K(\alpha) \ln K(\alpha) d\alpha + \frac{1}{2}\ln\left(\frac{2N}{\pi e}\right).
\end{equation}
And then subtracting $(1/2)\ln N$ as in Eq.~(\ref{newtilde}) gives us
\begin{equation}  \label{final}
\tilde{I} = - \int_0^{\pi/2} K(\alpha) \ln K(\alpha) d\alpha + \frac{1}{2}\ln\left(\frac{2}{\pi e}\right).
\end{equation}
The equality holds as long as our approximations become arbitrarily good as $N$ gets larger.  This will indeed be the case if
the function $K(\alpha)$ is reasonably well behaved.  A sufficient set of conditions on $K(\alpha)$ is that it be positive
and differentiable on the interval $[0,\pi/2]$.  Then when many trials are run, the range
of likely values of $\alpha$ narrows to such a degree that the final distribution $P(\alpha| n)$ does not depend appreciably on 
the {\em a priori} distribution $K(\alpha)$.  

The problem has now been reduced to finding out what distribution or distributions $K(\alpha)$ maximize the quantity $- \int_0^{\pi/2} K(\alpha) \ln K(\alpha) d\alpha$.  The answer to this question is well known: the unique maximizing distribution is the {\em uniform} distribution $K(\alpha) = 2/\pi$.  This result follows from the fact that the function $\phi(x) = -x\ln x$ is
a strictly concave function of $x$ for all positive values of $x$.  Jensen's inequality then tells us that
\begin{equation}
(2/\pi)\int_0^{\pi/2}\phi[K(\alpha)]d\alpha \le \phi\left[(2/\pi)\int_0^{\pi/2}K(\alpha) d\alpha\right] = \phi\left(2/\pi \right), 
\end{equation}
with equality holding only for the constant function $K(\alpha) = 2/\pi$ \footnote{Alternatively, instead of using differential entropies as in Eq.~(\ref{Kinfo}), we could have expressed the mutual information $I(\alpha : n)$ in terms of the Kullback-Leibler distances of both $K(\alpha)$ and 
$P(\alpha |n)$ from the uniform distribution over $\alpha$.  The calculation in Section III then tells us that $\tilde{I}$ is maximized when the 
Kullback-Leibler distance between $K(\alpha)$ and the uniform distribution is minimized, that is, when $K(\alpha)$ is itself the uniform 
distribution.}.  

Now we compare our result to quantum mechanics.  Is this uniform distribution over $\alpha$ the one induced by the quantum probability law
$p_0(\theta) = \cos^2 \theta$, when $\theta$ is uniformly distributed?  First consider the values of $\theta$ from 0 to $\pi/2$.  In that range
the law $p_0(\theta) = \cos^2 \theta$ mirrors the definition of $\alpha$ and we have $\alpha = \theta$.  (I am taking $p_1$ to correspond to 
the horizontal outcome.)  So a uniform distribution of $\theta$
over this range would induce the uniform distribution of $\alpha$.  In the other three quadrants of the circle, that is, in the rest of the
range of $\theta$, the parameter $\alpha$ is not equal to $\theta$ but we still have $|d\alpha/d\theta| = 1$ (except at a finite number of points where 
$\alpha$ ``bounces'' off one of the endpoints of its range).  Thus when $\theta$ is uniformly distributed, so is $\alpha$.  This completes our
demonstration that the quantum probability function $p_0(\theta) = \cos^2 \theta$ is optimal.  

Is the function $p_0(\theta) = \cos^2\theta$ unique in this respect?  The answer is no.  Any function $p(\theta)$ that yields the same {\em a priori}
measure on the binary probability space will be equally good.  For example, any function of the form $p(\theta) = \cos^2 (m\theta/2)$ where $m$ is
a non-negative integer yields the same distribution $K(\alpha) = 2/\pi$.  And there are many other, less physically interesting examples.  
Still, a typical function $p(\theta)$ will not have this optimization property.  

Looking back over the above argument, one can see that the crucial feature is the exponent in Eq.~(\ref{14}): the coefficient of $(\Delta \alpha)^2$ depends
only on $N$ and not on $\alpha$ itself.  In other words, the spread in the value of $\alpha^{(n)}$
depends only on the number of trials (when this number is large), and not on the probabilities $(p_1,p_2)$.  This is what is special about parameterizing
probability space with the parameter $\alpha$: it makes the statistical spread uniform.  Once we have this fact, it is guaranteed that the final differential
entropy $h(\alpha | n)$ will not depend
on $K(\alpha)$.  Therefore to maximize the mutual information, we want to maximize the initial differential entropy $h(\alpha)$ and we are thereby led to the uniform distribution.

\begin{figure}[h]
\begin{center}
\includegraphics[scale = 0.85]{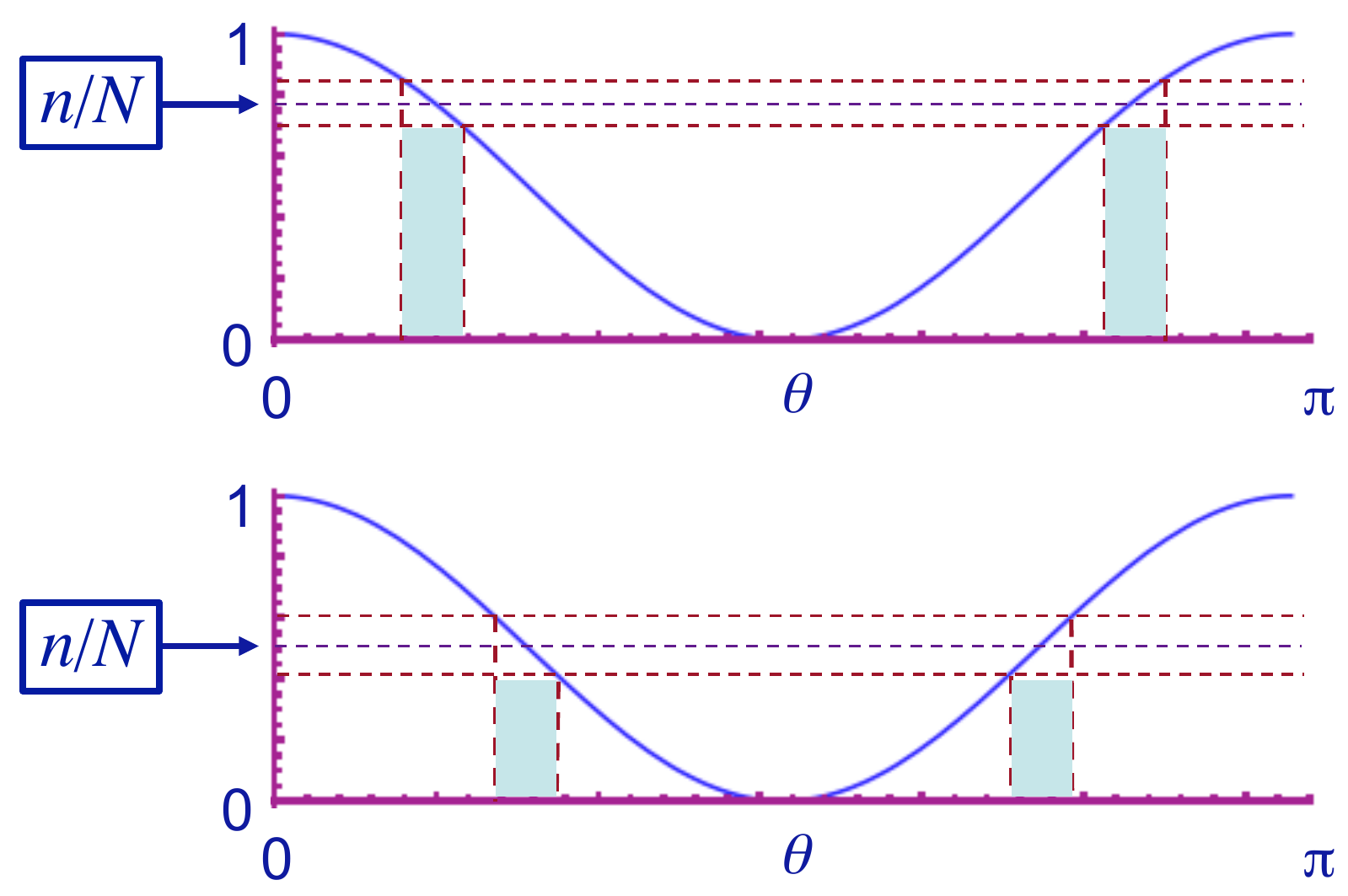}
\end{center}
\vspace{-10mm}
\caption{The uncertainty in $\theta$ for two different values of $n/N$.  Notice that the slope of the cosine-squared curve
exactly compensates for the varying size of $\Delta(n/N)$, so that the ``region of uncertainty'' in $\theta$ has the same
size for all values of $n/N$.  (Here $\theta$ is plotted only up to $\pi$ to make the diagram simpler.)}
\label{newequalizingfigure}
\end{figure}

There is perhaps a more direct way of seeing what is special about the function $p_0(\theta) = \cos^2\theta$.  First note that the spread in $n$ itself is {\em not} uniform over probability space.  If one performs the binary experiment $N$ times, the standard deviation in $n/N$ is
given by
\begin{equation}  \label{deltan}
\Delta (n/N) =\sqrt{\frac{p_1p_2}{N}},
\end{equation}
which is smaller near the ends of probability space than near the middle.  One can see this dependence in the exponent 
in Eq.~(\ref{gaussian}).  In our polarization experiment, an
observer recording the frequency of occurrence of ``horizontal'' will therefore be more certain of the {\em probability} of ``horizontal'' when that probability is close to 
zero or one.  (Again I am assuming that the experimenter's {\em a priori} distribution over probability space is reasonably smooth and that the number of trials is large.)  On the other hand, upon translating the uncertainty in probability to an uncertainty in $\theta$, the observer must use the
function $p(\theta)$.  For the special case of $p_0(\theta) = \cos^2\theta$, the slope of this function exactly compensates for the varying size
of $\Delta(n/N)$, so that the size of the resulting ``region of uncertainty'' of $\theta$ is independent of the value of $n/N$.  
Specifically,
\begin{equation}
\left| \frac{d}{d\theta} \cos^2\theta \right|= 2\left| \cos\theta\sin\theta \right| = 2\sqrt{p_0(\theta)[1-p_0(\theta)]},
\end{equation}
which perfectly matches the dependence seen in Eq.~(\ref{deltan}).
This compensation is illustrated in
Fig.~\ref{newequalizingfigure}.  Thus the Born rule has the effect of equalizing the final uncertainty in $\theta$ over all values
of $\theta$.  It is plausible that this even-handed strategy will be optimal, and indeed we have just seen that it is.   

We now consider the case in which {\em all} pure polarization states are possible.   The full set of pure states is the Bloch sphere---it includes the circular and elliptical polarizations---and the natural {\em a priori} measure is
the uniform measure on the sphere, since this is the only probability measure invariant under all unitary transformations.  We imagine 
a device that prepares a beam of photons in one of these polarization states, and further along the photons' path we imagine a person making the horizontal-vs-vertical measurement
on each photon.  The polarization is now determined by two parameters; for definiteness let us take them to be the polar angle $\beta$
and the azimuthal angle $\phi$, and let the north and south poles of the sphere correspond to horizontal and vertical polarization.  It is still possible to define the mutual information between the photons' preparation and the measurement
outcomes; it could be written as $I(n: \beta,\phi)$.  Again this mutual information is the same as the quantity $I(\alpha: n)$ given in
Eq.~(\ref{Kinfo}) and it is maximized only if the {\em a priori} distribution of $\alpha$ is the uniform distribution $K(\alpha) = 2/\pi$.   
But now the quantum mechanical law does {\em not} yield this distribution over the values of $\alpha$.  With the uniform distribution over the Bloch
sphere, the parameter $\cos\beta$ is uniformly distributed over the interval $[-1,1]$, and the quantum mechanical probability of ``horizontal,''
$p(\theta) = (1/2)(1 + \cos\beta)$, is therefore uniformly distributed over the interval $[0,1]$.  To get the corresponding distribution of $\alpha$, we use
the relation $p_1 = \cos^2\alpha$ and the assumption that $p_1$ is uniformly distributed:
\begin{equation}
K(\alpha) = \left| \frac{dp_1}{d\alpha}\right| = 2\cos\alpha\sin\alpha.
\end{equation}
Thus, rather than giving us the uniform distribution of $\alpha$, the full Bloch sphere gives us a distribution that has a maximum in the middle of $\alpha$'s range.  

We can see directly that this distribution does not allow as much information transfer as the optimal distribution.  The relevant quantity is the integral 
in Eq.~(\ref{final}):
\begin{equation}
- \int_0^{\pi/2} K(\alpha) \ln K(\alpha) d\alpha .
\end{equation}
For the uniform distribution over $\alpha$, this quantity has the value $\ln(\pi/2) = 0.452$, whereas for the distribution $K(\alpha) = 2\cos\alpha\sin\alpha$, 
we get $1-\ln2 = \ln(e/2) = 0.307$.

Just as the uniform measure over the surface of the Bloch sphere is natural because it is invariant under all unitaries, in the real-vector-space theory where the set of pure states in two dimensions traces out a circle rather than a sphere, the uniform distribution over the circle is natural because it
is invariant under all orthogonal transformations (rotations and reflections).  That is, in the real-vector-space theory, we can use this
invariance to justify our original assumption that the angle $\theta$ is uniformly distributed over the interval $[0,2\pi]$.

\section{Optimal Transfer of Information: The $d$-Dimensional Case}

The above argument extends to a $d$-dimensional real vector space.  Let a ``redit'' be a hypothetical quantum
object whose pure states are vectors in a $d$-dimensional vector space over the real numbers.  We now imagine an
experiment in which a beam of $N$ redits is prepared in a specific pure state $|s\rangle$.  At some point further along the beam, an observer
makes a fixed complete orthogonal measurement on each redit.  The observer records the integers $n_1, \ldots, n_d$, where $n_i$ is the
number of times the $i$th measurement outcome occurs.  We ask how much information the observer learns on average about the 
preparation $|s\rangle$, assuming (crucially) that the vector $|s\rangle$ is initially distributed uniformly over the unit sphere in the $d$-dimensional
space.  Again this average information gain is given by the mutual information, which we will write down shortly.  The mutual information
depends on the law that specifies the probability of the $i$th outcome given the preparation $|s\rangle$.  In real-vector-space quantum
theory, this law can be expressed as 
\begin{equation}  \label{Bornrule}
{p}_i(|s\rangle) = s_i^2,  \hspace{1cm} i = 1, \ldots, d,
\end{equation}
where $s_1, \ldots, s_d$ are the components of $|s\rangle$ in the basis defined by the measurement.  

As before, what really matters in computing the mutual information is the {\em a priori} measure on probability space.  The uniform
measure over the unit sphere in $d$ dimensions, together with Eq.~(\ref{Bornrule}), defines some specific {\em a priori} measure on
probability space.  We also want to consider other {\em a priori} measures, in order to show that the one induced by
Eq.~(\ref{Bornrule}) is optimal.  The probability space is now a $(d-1)$-dimensional set, since the probabilities must add to one.
We could parameterize this set by the probabilities $p_1, \ldots, p_{d-1}$ of the first $d-1$ outcomes, but we instead choose to label the points of probability space by a unit vector $\vec{\gamma} = 
(\sqrt{p_1}, \ldots, \sqrt{p_d})$.  (Note that each $\sqrt{p_i}$ is non-negative; so $\vec{\gamma}$ is confined to
the positive part of the unit sphere.)  We could go further and choose $d-1$ specific angular coordinates to locate this vector
on the sphere (like the $\alpha$ of the preceding section), but we will not need to do so.  Let $K(\vec{\gamma})d\vec{\gamma}$ be a generic {\em a priori} probability measure on the set of
vectors $\vec{\gamma}$, where $d\vec{\gamma}$ is an infinitesimal $(d-1)$-dimensional surface element on the positive section of the unit sphere.  Our goal is to find the distribution $K(\vec{\gamma})$ that maximizes the mutual information.  

That mutual information can be written as follows:
\begin{equation}  \label{Imulti}
I(\vec{\gamma}:\vec{n}) = h(\vec{\gamma}) - h(\vec{\gamma}|\vec{n}) = 
-\int K(\vec{\gamma})\ln K(\vec{\gamma}) d\vec{\gamma} + \sum P(\vec{n})\int P(\vec{\gamma}|\vec{n})\ln P(\vec{\gamma}|\vec{n}) d\vec{\gamma},
\end{equation}
where $\vec{n} = (n_1, \ldots, n_d)$ specifies the number of times each outcome occurs.  The sum is over all vectors
$\vec{n}$ for which each $n_i$ is a non-negative integer and $n_1 + \cdots + n_d = N$.  The mutual information is now based on the multinomial distribution:
\begin{equation} \label{multinomial}
P(\vec{n}|\vec{\gamma}) = \frac{N!}{n_1! \cdots n_d!} p_1^{n_1} \cdots p_d^{n_d}.
\end{equation}
(Here $p_j = \gamma_j^2$.)
The functions appearing in Eq.~(\ref{Imulti}) can be obtained from $P(\vec{n}|\vec{\gamma})$ as follows:
\begin{equation}  \label{Pn}
P(\vec{n}) = \int P(\vec{n}|\vec{\gamma}) K(\vec{\gamma}) d\vec{\gamma}
\end{equation}
and
\begin{equation}  \label{Pgamman}
P(\vec{\gamma}|\vec{n}) = \frac{P(\vec{n}|\vec{\gamma})K(\vec{\gamma})}{P(\vec{n})}.
\end{equation}
As $N$ gets large, it will turn out that $I(\vec{\gamma} : \vec{n})$ grows as $[(d-1)/2]\ln N$.  So we will compute the limiting value
\begin{equation} \label{multitilde}
\tilde{I} = \lim_{n\rightarrow\infty} \left[I(\vec{\gamma} : \vec{n}) - \left( \frac{d-1}{2}\right) \ln N\right].
\end{equation}
At this point the calculation is very similar to the one 
in the preceding section.  As we did for the analagous equation in that case, we now show that the second term on the right-hand side of Eq.~(\ref{Imulti}) becomes independent of $K(\vec{\gamma})$
as $N$ approaches infinity.  

For any fixed positive values of $p_1, \ldots, p_d$ and 
for large enough $N$,
Eq.~(\ref{multinomial}) can be approximated arbitrarily well by a Gaussian function:
\begin{equation}  \label{multigaussian}
P(\vec{n}|\vec{\gamma}) \approx \left[(2\pi N)^{d-1} p_1p_2 \cdots p_d\right]^{-1/2}\exp\left[-\frac{N}{2}\sum_{i=1}^d \frac{(n_i/N - p_i)^2}{p_i}\right].
\end{equation}
It will be helpful to define the vectors
\begin{equation}
\vec{\gamma}^{(n)} = \left( \sqrt{\frac{n_1}{N}}, \ldots, \sqrt{\frac{n_{d}}{N}}\right) \hspace{1cm} \hbox{and}\hspace{1cm} \vec{\Delta\gamma} = 
\vec{\gamma}^{(n)} - \vec{\gamma}.
\end{equation}
That is, $\vec{\gamma}^{(n)}$ is the vector of square roots of the observed frequencies of occurrence, whereas $\vec{\gamma}$ is 
the vector of square roots of the probabilities.  The difference $\vec{\Delta\gamma}$ between these two vectors is likely to be small when $N$ is large;  
so we will keep just the lowest-order term
in this quantity.  We can then rewrite the sum in the 
exponent of Eq.~(\ref{multigaussian}) as follows:
\begin{equation}
\sum_{i=1}^d \frac{\left(n_i/N - p_i\right)^2}{p_i}
= \sum_{i=1}^d \frac{\left(\Delta p_i \right)^2}{p_i} \approx 4\left|\vec{\Delta\gamma}\right|^2,
\end{equation}
where $\Delta{p_i}$ is defined to be $(n_i/N) - p_i$ and the last step comes from
\begin{equation}
\Delta \gamma_i = \Delta\left( p_i^{1/2}\right) \approx \frac{1}{2}p_i^{-1/2}\Delta p_i.
\end{equation}
We can therefore approximate Eq.~(\ref{multigaussian}) as
\begin{equation}  \label{simplegaussian}
P(\vec{n}|\vec{\gamma}) \approx (2\pi N)^{-\left(\frac{d-1}{2}\right)}\frac{1}{\gamma_1\gamma_2 \cdots \gamma_d}
\exp\left( - 2N\left|\vec{\Delta\gamma}\right|^2 \right).
\end{equation}
That is, $P(\vec{n}|\vec{\gamma})$, regarded as function of $\vec{\gamma}$, falls off as a Gaussian around the 
point $\vec{\gamma}^{(n)}$, with a spread that is isotropic and independent of the value of $\vec{\gamma}^{(n)}$.  (This function is not, however,
a {\em normalized} distribution of $\vec{\gamma}$; rather, it is normalized with respect to a sum over $\vec{n}$.)

In approximating the integral in Eq.~(\ref{Pn}), we rely on the narrowness of the Gaussian: the integral is over a section of a sphere, but we can treat it
as being over an infinite flat space having $d-1$ dimensions---the ``plane'' tangent to the sphere at the point $\vec{\gamma}^{(n)}$.  We also evaluate
everything outside the exponential at the point $\vec{\gamma}^{(n)}$.  These approximations give us
\begin{equation}
P(\vec{n}) \approx (2N)^{-(d-1)}\frac{K(\vec{\gamma}^{(n)})}{\gamma_1^{(n)}\gamma_2^{(n)}\cdots\gamma_d^{(n)}}.
\end{equation}
Inserting this expression and Eq.~(\ref{simplegaussian}) into Eq.~(\ref{Pgamman}), we get
\begin{equation}
P(\vec{\gamma}|\vec{n}) \approx \left( \frac{2N}{\pi}\right)^{\frac{d-1}{2}} \exp\left( - 2N\left| \vec{\Delta\gamma} \right|^2 \right).
\end{equation}
We can now do the second integral in Eq.~(\ref{Imulti}), again treating the integral as if it were over an infinite $(d-1)$-dimensional flat space.  
The result is 
\begin{equation}
I(\vec{\gamma}: \vec{n}) \approx -\int K(\vec{\gamma})\ln K(\vec{\gamma}) d\vec{\gamma} + \left(\frac{d-1}{2}\right) \ln\left( \frac{2N}{\pi e}\right).
\end{equation}
Finally we subtract $[(d-1)/2]\ln N$ as in Eq.~(\ref{multitilde}) to get
\begin{equation}
\tilde{I} = -\int K(\vec{\gamma})\ln K(\vec{\gamma}) d\vec{\gamma} + \left(\frac{d-1}{2}\right) \ln\left( \frac{2}{\pi e}\right).
\end{equation}
Note that Eq.~(\ref{final}) is a special case of this equation, with $d=2$.  As in that case, the expression is maximized
by choosing $K(\vec{\gamma})$ to correspond to the uniform distribution:
\begin{equation}  \label{multiK}
K_{opt}(\vec{\gamma}) =   
\frac{2^d\, \Gamma\hspace{-1mm}\left(\frac{d}{2}+1\right)}{d\, \pi^{d/2}}.
\end{equation}
The constant on the right-hand side of Eq.~(\ref{multiK}) is the reciprocal of the ``surface area'' of the positive section of the unit sphere.  

The question now is whether the probability rule in real-vector-space quantum theory, $p_i = s_i^2$, induces the measure 
on probability space given by $K_{opt}(\vec{\gamma})$.  The answer is yes, as is easily seen.  The state vector $|s\rangle$
ranges over the full unit sphere in ${\mathbb R}^d$, but consider for now just the section of the sphere in which each $s_i$
is positive.  In that section the vector $\vec{\gamma}$ is equal to the vector $|s\rangle$, since
$p_i = \gamma_i^2 = s_i^2$.  So the uniform distribution of $|s\rangle$ over this section of the sphere induces the uniform
distribution of $\vec{\gamma}$.  The whole unit sphere in ${\mathbb R}^d$ consists, in effect, of $2^d$ copies of the 
positive section.  So indeed, the uniform distribution of $|s\rangle$ over the whole sphere does correspond to the 
uniform distribution of $\vec{\gamma}$ expressed in Eq.~(\ref{multiK}).  That is, the transfer of information from preparation to measurement
is optimized in this $d$-dimensional case, just as it was in the two-dimensional case.  

\begin{figure}[h]
\begin{center}
\includegraphics[scale = 0.8]{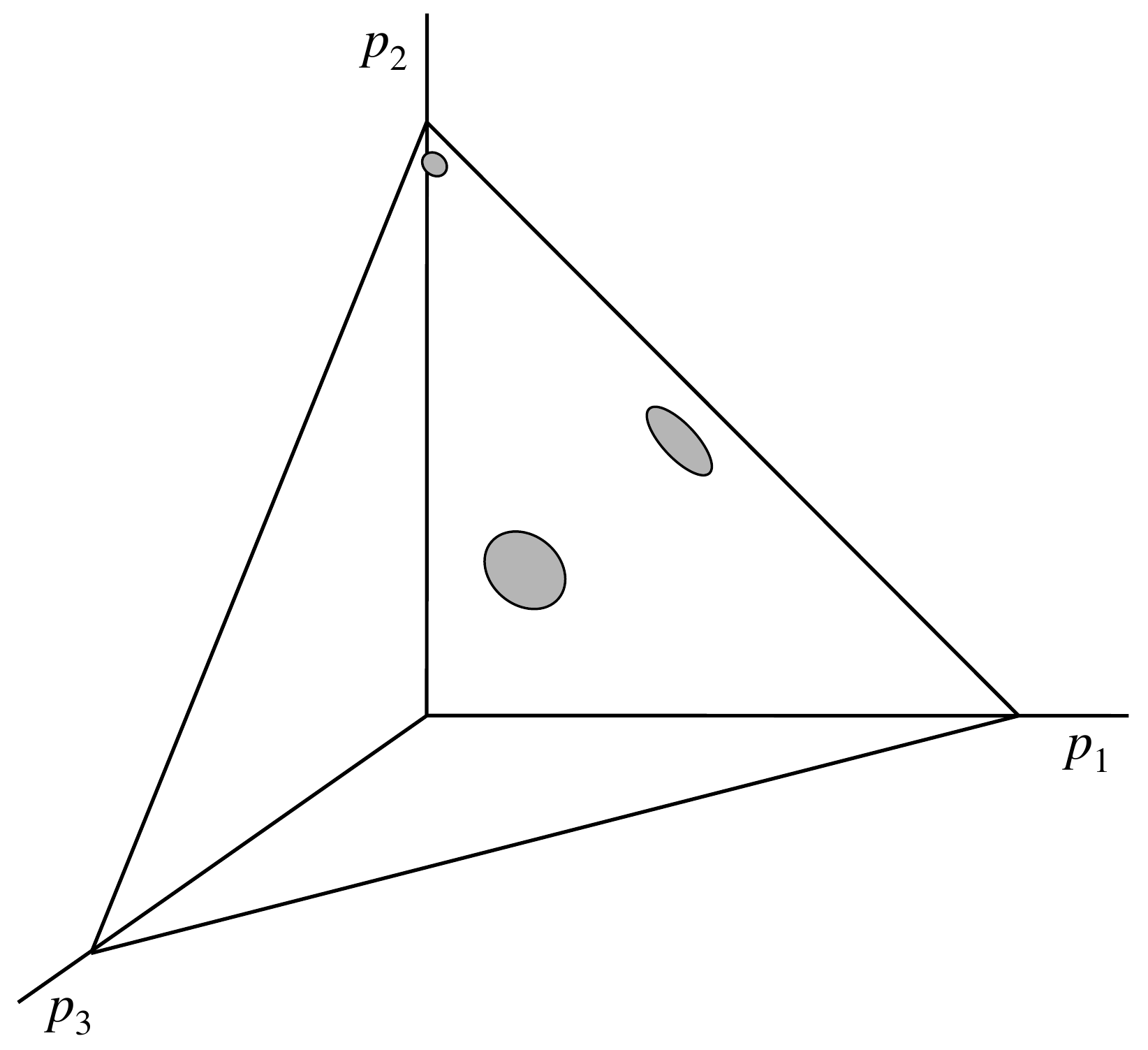}
\end{center}
\vspace{-10mm}
\caption{Regions of uncertainty at different locations in the flat probability space.  As one approaches an edge, the uncertainty shrinks along the
direction perpendicular to the edge.}
\label{probspace}
\end{figure}

\begin{figure}[h]
\begin{center}
\includegraphics[scale = 0.8]{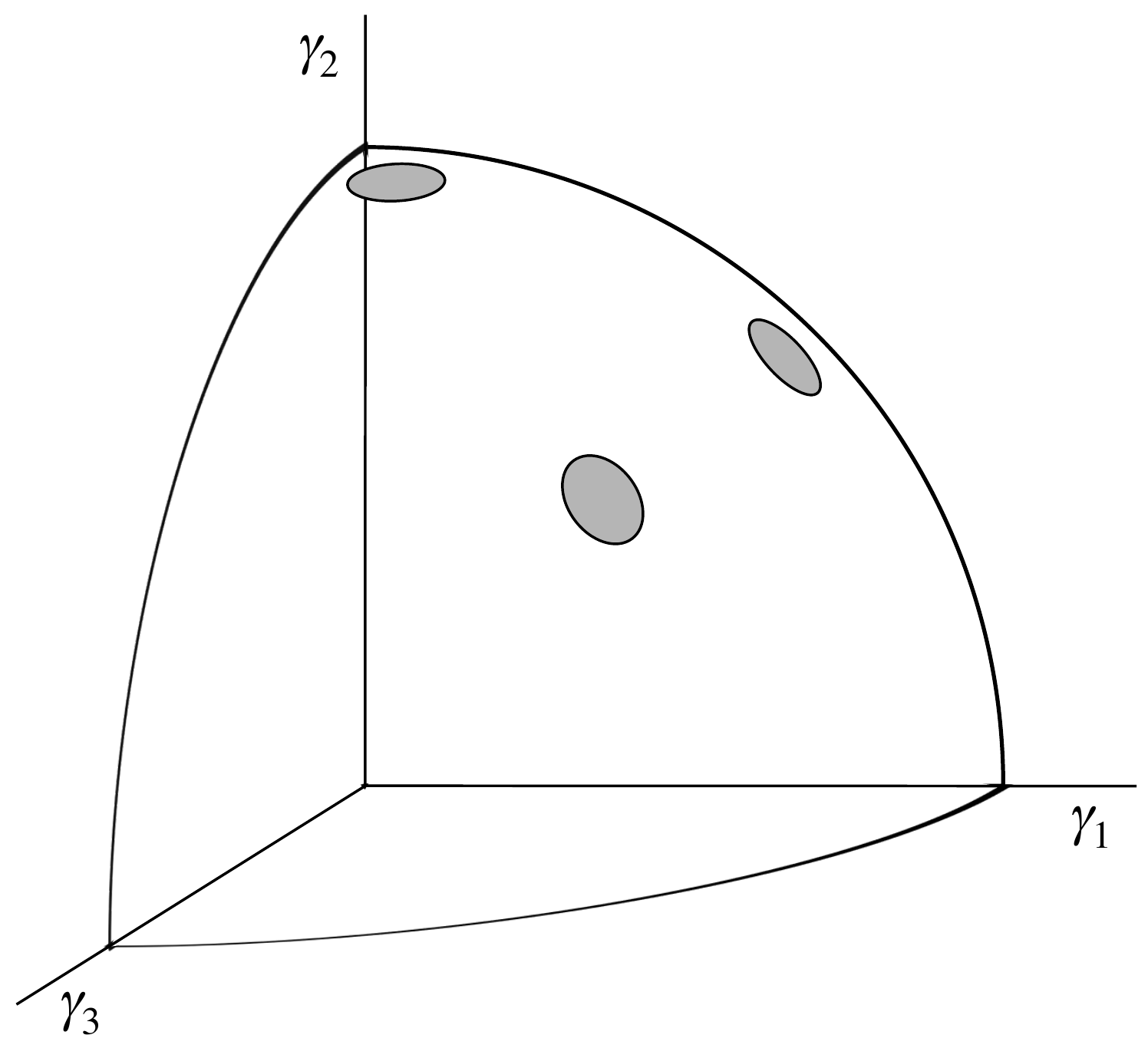}
\end{center}
\vspace{-10mm}
\caption{Regions of uncertainty at different locations in the spherical probability space, with axes corresponding to the square roots 
of probability.  Now all regions of uncertainty are isotropic and of the same size.}
\label{spherical}
\end{figure}

The complications involved in the information-theoretic calculation may obscure what is really a simple underlying fact.  
Imagine probability space as a $(d-1)$-dimensional flat ``surface'' in a $d$-dimensional space with orthogonal axes
labeled $p_1, \ldots, p_d$.  The surface consists of all points $(p_1, \ldots, p_d)$ such that each $p_i$ is non-negative
and the sum $p_1 + \cdots + p_d$ is equal to 1.  Around each point on this surface one can imagine a small ``region of
uncertainty,'' representing the spread in the actual frequencies of occurrence of the $d$ possible outcomes in $N$
trials.  These regions of uncertainty can be derived from the multinomial distribution, Eq.~(\ref{multinomial}),
and for large $N$ their sizes and shapes can be read off the exponent in Eq.~(\ref{multigaussian}).  (We could, for example,
define ``region of uncertainty'' to be the range of values of $(n_1/N, \ldots, n_d/N)$ for which the exponent has magnitude
less than 1.)  One can see that these regions of uncertainty will have different sizes and shapes, depending on the location 
in probability space.  For example, the largest is at the exact center, as shown in Fig.~\ref{probspace}.  
But if we change the axes of probability space from
$p_i$ to $\gamma_i = \sqrt{p_i}$, probability space then looks like a section of a sphere.  Again one can speak of a
region of uncertainty around each point on this spherical surface, but now it happens that all the regions of uncertainty
have the same size and shape---in fact they are all spherical---as we can see in the exponent of Eq.~(\ref{simplegaussian}) and as is illustrated
in Fig.~\ref{spherical}.  (The issue gets tricky near
the edges.  The closer one gets to the edge, the higher the value of $N$ must be in order to see this uniformity.  But 
no matter how close one is to the edge---as long as one is not {\em at} the edge---there is always such a value 
of $N$.)  In this sense, there is
something special about representing probability space as a section of a sphere: it captures geometrically the statistical fluctuations
in a large sample.   What is special about real-vector-space quantum theory is that its set of pure states mirrors this representation of probability space.    

As one would expect, the mathematical fact illustrated in Fig.~\ref{spherical} has been well noted in the statistics literature.  Bhattacharyya in the 1940's 
proposed a distance measure between two probability distributions based on the {\em angle} between their $\vec{\gamma}$
vectors \cite{Bhattacharyya}.  The square-root construction has been particularly explicit in the genetics literature.  One can see diagrams 
of the positive section of the unit sphere in papers by Cavalli-Sforza and collaborators from the 1960's, and these authors
give credit for the idea to R.~A.~Fisher \cite{Cavalli1, Cavalli2} (as do Mosteller and Tukey \cite{Mosteller}).  In the present paper, I have used the square-root construction only to identify a special {\em measure} on probability space---the uniform measure on the spherical surface traced out by $\vec{\gamma}$.  But one can also use it
to define a special {\em metric} on the space, and this is what Bhattacharyya, Cavalli-Sforza and others have done.  (One can find in Ref.~\cite{Takezaki}
a review a various ``genetic distances,'' some of which are based on the square-root construction.)  Such a metric has also
been used in work on quantum foundations \cite{Wootters2, Wootters3, Braunstein, Goyal}.  However, I want to emphasize that this special feature of the representation of probability space in terms of square roots of probability
arises without any reference to quantum theory.  It is simply a matter of statistics.  

What about ordinary complex-vector-space quantum theory?  In that theory each pure state is represented by a vector
$|s\rangle$ in ${\mathbb C}^d$.  The natural {\em a priori} distribution over pure states is the uniform distribution
over the unit sphere in ${\mathbb C}^d$, that is, the unique distribution invariant under all unitary transformations. (We could just 
as well speak of a distribution over projection operators $|s\rangle\langle s|$ so as not to have to worry about the
irrelevant overall phase factor in the vector $|s\rangle$, but for our purposes either picture leads to the same result.)  
For a complete orthogonal measurement, the probabilities of the outcomes are given by $p_i = |s_i|^2$, where the $s_i$'s
are the components of $|s\rangle$ in the basis defined by the measurement.  We can ask what measure this probability rule, together
with the {\em a priori} distribution of state vectors $|s\rangle$, induces on probability space.  That question was answered
by Sykora in 1974: it induces the {\em uniform} distribution, not on the spherical surface defined by $\vec{\gamma}$, but
on the flat surface defined by $(p_1, \ldots, p_d)$ \cite{Sykora}.  This is a remarkably simple and intriguing result, but this distribution is not
the one that optimizes the transfer of information from preparation to measurement.  

\section{Optimal Information Transfer in Standard Quantum Mechanics?}

The real-vector-space theory thus has a certain elegance to it, in that there is an optimal correspondence between the set of pure states
and the set of probability distributions over the outcomes of a complete orthogonal measurement.  The complex theory does not have this
property, but one might wonder whether this is because we are not asking the question in the right way.  That is, by somehow reframing the problem,
might it be possible to see that the usual complex theory does exhibit the property of optimal information transfer in some altered sense?  

For example, perhaps we are making a mistake to consider a complete orthogonal measurement.  Such a measurement will never 
reveal the relative phases between the components of the state vector when it is written in the measurement basis.  Instead we could consider a special case of a non-orthogonal measurement, namely, a symmetric informationally-complete measurement (a SIC) \cite{Zauner, Caves, Renes}.  In $d$ complex dimensions,
such a measurement has $d^2$ possible outcomes.  Each outcome corresponds to a pure state $|m_i\rangle\langle m_i|$, and the 
inner product between any two of these pure states has the same magnitude: $|\langle m_i|m_j\rangle |^2 = 1/(d+1)$.  Numerical evidence strongly indicates that
such symmetric measurements exist for all values of $d$ up to 67 \cite{Scott}, and it would be reasonable to guess that they exist for all $d$.  Such symmetric
measurements figure prominently in the quantum Bayesian approach to understanding quantum mechanics \cite{Appleby, Fuchs}.
Is there a kind of optimal transfer of information from preparation to measurement that occurs when the measurement is a SIC?

With $d^2$ possible outcomes, one can estimate $d^2 - 1$ independent parameters by repeating the measurement on many identically prepared
copies.  This is exactly the number of parameters needed to specify a $d \times d$ density matrix, and indeed, any density matrix can be reconstructed
with arbitrary precision from a fixed SIC applied to many copies.  (This is the meaning of ``informationally complete.'')  To state the question of optimal
information transfer, we would need to specify an {\em a priori} measure on the set of all $d\times d$ density matrices.  The measure should be 
unitarily invariant, but there are many unitarily invariant measures on this set.  Is there at least one such measure for which 
the mutual information between the preparation (of a general mixed state) and the measurement outcomes is optimal?  

One can see quickly that there is no such measure.  The optimal {\em a priori} measure on probability space has already been determined in the
preceding section.  For $d^2$ possible outcomes, the optimal measure is the uniform measure over the $(d^2 - 1)$-dimensional spherical surface of probability
space, when the axes correspond to the square roots of the probabilities.  This measure clearly assigns nonzero weight to every nonzero volume
of probability space.  But if one
performs a SIC on any state, the largest possible value of any probability is $1/d$.  Thus the SIC does not make use of the whole probability
space; so it is not providing information optimally in our sense, no matter what weighting function we place on the set of density matrices.  

Let us try another version of the problem.  Suppose we are given a specific entangled state of a pair of qubits, namely, the state
\begin{equation}
|\Phi^+\rangle = \frac{1}{\sqrt{2}}\left(|00\rangle + |11\rangle \right).
\end{equation}
We imagine the first qubit is held by Alice and the second by Bob.  Now Alice applies a unit-determinant unitary transformation $U$ to her qubit---an element of $SU(2)$.  She then sends her
qubit to Bob, who performs a Bell measurement on the two qubits.  That is, he distinguishes the four orthogonal states
\begin{equation}  \label{Bell}
\begin{split}
|\Phi^+\rangle = \frac{1}{\sqrt{2}}\left(|00\rangle + |11\rangle \right) \hspace{2cm}
|\Phi^-\rangle = \frac{1}{\sqrt{2}}\left(|00\rangle - |11\rangle \right)  \\
|\Psi^+\rangle = \frac{1}{\sqrt{2}}\left(|01\rangle + |10\rangle \right) \hspace{2cm}
|\Psi^-\rangle = \frac{1}{\sqrt{2}}\left(|01\rangle - |10\rangle \right)
\end{split}
\end{equation}
We imagine this whole procedure is repeated over and over---always with the same initial state, the same $U$, and the same Bell measurement---so
that Bob can try to gain information about $U$ from the outcomes of his measurements.  We assume he already knows the initial state $|\Phi^+\rangle$.  (This scenario is like superdense coding \cite{Bennett}, except that we are not restricting $U$ to a discrete set.  Really what Bob is doing here is a restricted kind of process tomography \cite{Chuang, Poyatos}---trying to infer the process $U$ from the outcomes of measurements.) We can ask whether the transfer of information
is optimal between Alice's choice of unitary transformation and the outcomes of Bob's measurements.  

A general element of $SU(2)$ can be represented as
\begin{equation}
U = \exp\left[i(\theta/2)\hat{n}\cdot \vec{\sigma}\right],
\end{equation}
where $\hat{n}$ is the unit vector defining the axis of the Bloch sphere around which Alice is rotating her qubit, $\theta$ is the angle of rotation---it runs from
zero to $2\pi$---and $\vec{\sigma}$ is the vector of Pauli matrices.  The transformations $U$ are in one-to-one correspondence with the points of 
a three-dimensional spherical surface, which we can imagine embedded in four dimensions.  Specifically, we can label the point corresponding to the above $U$ 
by the unit vector
\begin{equation}  \label{suvec}
v_U = (\cos(\theta/2), n_x \sin(\theta/2), n_y \sin(\theta/2), n_z \sin(\theta/2)).
\end{equation}  
The natural measure on $SU(2)$ is the uniform
measure over this sphere---it is the unique measure that is invariant under left-multiplication (or right-multiplication) by any group element.  

To determine whether the information is transferred optimally from Alice to Bob, we need to compute the probabilities of Bob's outcomes.  It is 
straightforward to do so, and one finds that the probabilities are, in an arrangement parallel to that given in Eq.~(\ref{Bell}), 
\begin{equation}
\begin{split}
&\cos^2(\theta/2) \hspace{2.4cm} n_z^2 \sin^2(\theta/2) \\
&n_x^2 \sin^2(\theta/2) \hspace{2cm} n_y^2 \sin^2(\theta/2)
\end{split}
\end{equation}
These probabilities are the squared components of the unit vector $v_U$ given in Eq.~(\ref{suvec}).  Thus the problem is equivalent to
the case of real-vector-space quantum mechanics in four dimensions.  So indeed, the information is transmitted optimally 
from Alice to Bob!  

Does this example generalize to higher dimensions?  The answer is no, at least not in any way that I can see.  For example, in three dimensions,
we would probably want Alice and Bob to start with the state $|\Phi\rangle = (|00\rangle + |11\rangle + |22\rangle)/\sqrt{3}$.  Alice will perform a general
unit-determinant unitary transformation $U$, and then Bob will measure both particles in the generalized Bell basis, which consists of the nine
states
\begin{equation}  \label{3bell}
\begin{split}
&|00\rangle + |11\rangle + |22\rangle \hspace{1.5cm} |00\rangle + \omega |11\rangle + \omega^2 |22\rangle 
\hspace{1.5cm} |00\rangle + \omega^2 |11\rangle + \omega |22\rangle \\
&|01\rangle + |12\rangle + |20\rangle \hspace{1.5cm} |01\rangle + \omega |12\rangle + \omega^2 |20\rangle 
\hspace{1.5cm} |01\rangle + \omega^2 |12\rangle + \omega |20\rangle \\
&|02\rangle + |10\rangle + |21\rangle \hspace{1.5cm} |02\rangle + \omega |10\rangle + \omega^2 |21\rangle 
\hspace{1.5cm} |02\rangle + \omega^2 |10\rangle + \omega |21\rangle.
\end{split}
\end{equation}
Here $\omega = e^{2\pi i/3}$ and I have suppressed the normalization factor $1/\sqrt{3}$.  A counting of parameters is initially encouraging:
it takes eight real numbers to specify an element of $SU(3)$, and Bob's measurement yields eight independent probabilities.  However,
one quickly discovers that, as in the case of the SIC, the measurement does not make use of the whole probability space.  

Consider specifically
the probabilities of the second and third outcomes listed on the first row of Eq.~(\ref{3bell}); let us call these probabilities $p_2$ and $p_3$ (we imagine
a list of nine probabilities $p_1, \ldots, p_9$, of which these are the second and third).  In terms of the components $u_{ij}$ of Alice's unitary matrix $U$,
we have
\begin{equation}
p_2 = \frac{1}{9}\left| u_{00} + \omega^2 u_{11} + \omega u_{22} \right|^2 \hspace{1cm}\hbox{and}\hspace{1cm}
p_3 = \frac{1}{9}\left| u_{00} + \omega u_{11} + \omega^2 u_{22} \right|^2,
\end{equation}
so that the product has the value
\begin{equation}  \label{usum}
p_2p_3 = \frac{1}{81} \left| u_{00}^2 + u_{11}^2 + u_{22}^2 - u_{00}u_{11} - u_{00}u_{22} - u_{11}u_{22} \right|^2.
\end{equation}
Now, in the whole probability space the maximum value of $p_2p_3$ is 1/4, attained when $p_2 = p_3 = 1/2$.  But given that
each $u_{ij}$ can have a magnitude no larger than 1, one can show that the expression
in Eq.~({\ref{usum}) cannot exceed the value $16/81 < 1/4$ \footnote{In proving this inequality, we are free to set $u_{00}$ equal to 1.
Then let $u_{11} = -a$ and $u_{22} = -b$ and the desired inequality becomes
$$
|1 + a + b + a^2 + b^2 - ab| \le 4
$$
under the assumption that $|a| \le 1$ and $|b| \le 1$.  (One can see that equality is achieved when $a = b = 1$.)  This inequality is equivalent to 
$$
\left|A^2 + B^2 + (A-B)^2 \right| \le 8,
$$
where $A = 1+a$ and $B = 1+b$.  This last inequality can be proved by first noting that
$$
\left|A^2 + B^2 + (A-B)^2 \right| \le |A^2 + B^2| + |A-B|^2
$$
and then writing out the absolute values explicitly.  One has to use the fact that $A$ and $B$ are both confined to a circle of unit radius in the complex plane,
centered at the value 1.}.  Thus a certain region of probability space is
inaccessible in the scenario we are imagining.  It follows that the information about $U$ is not conveyed optimally to Bob through his
measurement outcomes.  

Thus, as far as I can tell, the property of optimal information transfer does not easily carry over from the real-vector-space theory
to ordinary quantum mechanics.  

\section{Conclusions}

In real-vector-space quantum theory, the number of parameters needed to specify a pure state is equal to the number of 
independent probabilities in a complete orthogonal measurement: both are equal to $d-1$ for a state in $d$ dimensions.  So by measuring many copies 
prepared in an unknown pure state, 
one can hope to pin the state down to a finite number of small regions in state space.  In this paper we have seen
that this pinning down is in fact optimal, in the sense that the observer gains as much information about the state as could possibly be 
gained in any probabilistic theory, at least when the number of trials is very large.  

Standard quantum theory, based on a complex vector space, does not have this property, and we have not been able to find
a restatement of the problem for which the complex theory does achieve such an optimization (except for the case of a unitary transformation
applied to a qubit).  For our original statement of the problem, one can say that this lack of optimization comes from the fact that for any specific orthogonal measurement,
a complete specification of a pure state includes phase factors that have no effect on the probabilities of the outcomes.  The presence of these phase factors changes the 
natural {\em a priori} measure on probability space, and the mutual information is no longer maximized.  

Note that in the complex theory the number of real parameters needed to specify a pure state in $d$ dimensions is $2d - 2$ if we do not count an irrelevant overall phase factor.  This number is exactly {\em twice} the number of
independent probabilities an orthogonal measurement can access, and it seems that this doubling of the number of parameters is what spoils 
the optimization.  It is reasonable to ask whether there can be some deeper understanding of this factor of two, but at this point it is hard to have confidence in 
any particular answer.

In a sense, any axiomatization of quantum mechanics
offers a potential answer to this question: whatever assumptions give rise to the structure of quantum theory also give rise
to the factor of two.  In his axiomatization, Goyal addresses the factor of two directly, formalizing it in his principle of complementarity: for a measurement 
that at some level generates $2d$ possible events, only $d$ distinguishable outcomes can be observed, each corresponding to a {\em pair} of 
fundamental events.  This principle, together with a principle of global gauge invariance, 
leads him to the basic structure of quantum mechanics \cite{Goyal}.  One achieves a similar result by assuming that the underlying theory 
is real-vector-space quantum theory, but that because our knowledge is limited in some fundamental way we do not see the full real-vector-space structure; we have access only to those observables that satisfy Stueckelberg's rule, that is, those observables that commute with the operator $I \otimes J$ defined in the introduction.  (Goyal in fact relates his work to Stueckelberg's.  See also Ref.~\cite{Aleksandrova}, in which Stueckelberg's rule emerges dynamically.)  Imposing Stueckelberg's rule on a real vector space of $2d$ dimensions reduces the maximum number of
orthogonal states from $2d$ to $d$, and it cuts in half the number of parameters required to specify a maximally pure state \footnote{When standard
quantum mechanics is expressed in real-vector-space terms, what we normally call a pure state is represented by a density matrix of
rank 2.}.  

While such an interpretation would give an important role to the real-vector-space theory, it raises a difficult question about the status of the main result in this paper.  If the limitation on our knowledge
is fundamental, then who are the observers for whom the transfer of information from preparation to measurement is optimal?  Evidently it is not optimal for us, because 
whatever the underlying theory may be, the {\em effective} theory within which we live seems to be complex-vector-space quantum theory.

\end{document}